\documentclass[aps,10pt,twocolumn,superscriptaddress,footinbib,flushbottom]{revtex4-1}%

\usepackage{amssymb,amsmath,amsfonts,bm,dcolumn}
\usepackage[symbol]{footmisc}
\usepackage{graphicx,color}
\usepackage[squaren]{SIunits}
\usepackage[english]{babel}
\usepackage{tabularx}
\usepackage{tikz}
\usepackage{epstopdf}
\usepackage{placeins} 

\usepackage{times}

\begin{document} 

\title{Tuning electrostatic interactions of colloidal particles at oil-water interfaces with organic  salts} 
	
\author{Carolina van Baalen}
    \affiliation{Laboratory for Soft Materials and Interfaces, Department of Materials, ETH Z{\"u}rich, Vladimir-Prelog-Weg 5, 8093 Z{\"u}rich, Switzerland}
\author{Jacopo Vialetto}
    \altaffiliation{Current address: Department of Chemistry, University of Florence, via della Lastruccia 3, Sesto Fiorentino, I-50019 Firenze, Italy}
    \affiliation{Laboratory for Soft Materials and Interfaces, Department of Materials, ETH Z{\"u}rich, Vladimir-Prelog-Weg 5, 8093 Z{\"u}rich, Switzerland}	
\author{Lucio Isa}
    \email{E-mail: lucio.isa@mat.ethz.ch}
    \affiliation{Laboratory for Soft Materials and Interfaces, Department of Materials, ETH Z{\"u}rich, Vladimir-Prelog-Weg 5, 8093 Z{\"u}rich, Switzerland}

\begin{abstract} 
Monolayers of colloidal particles at oil-water interfaces readily crystalize owing to electrostatic repulsion, which is often mediated through the oil. However, little attempts exist to control it using oil-soluble electrolytes. We probe the interactions amongst charged hydrophobic micospheres confined at a water/hexadecane interface and show that repulsion can be continuously tuned over orders of magnitude upon introducing minor amounts of an organic salt into the oil. Our results show that charged groups at the particle/oil interface are subject to an associative discharging mechanism, analogous to the charge regulation kinetics observed for charged colloids in non-polar solvents. 
\end{abstract}
	
	\pacs{???}
	
	\maketitle


Particles adsorbed at a fluid interface are key components in multiple technological processes, from mineral recovery in froth flotation \cite{Tran2019}, to the fabrication of ordered two-dimensional (2D) materials \cite{Vogel2015a}. Confinement at an interface makes it also possible to obtain model systems, \textit{e.g.} to study crystallization in 2D \cite{Eisenmann2004a, Eisenmann2004b, vonGruenberg2004, Keim2004}. Regardless of the final goal, understanding and controlling the interactions between particles in interfacial monolayers is essential to achieve the desired structural and mechanical properties. To this end, a variety of additives \cite{Vialetto2021FA}, such as salts or surfactants \cite{Reynaert2006,Srivastava2014,Li-Destri2019,Reynaert2007,Truzzolillo2016,Anyfantakis2018,Vialetto2020}, and external stimuli (\textit{e.g.}, magnetic fields or light) \cite{Snezhko2011,Vialetto2019,Vialetto2021} have been used to tune particle assembly, organization and the mechanical stability of the resulting monolayers. While these phenomenological approaches result in a fine tuning of the monolayers' properties, unraveling the precise nature of the inter-particle potential still presents opportunities.

\begin{figure}[t!]
    \centering
    \includegraphics[width=0.9\columnwidth]{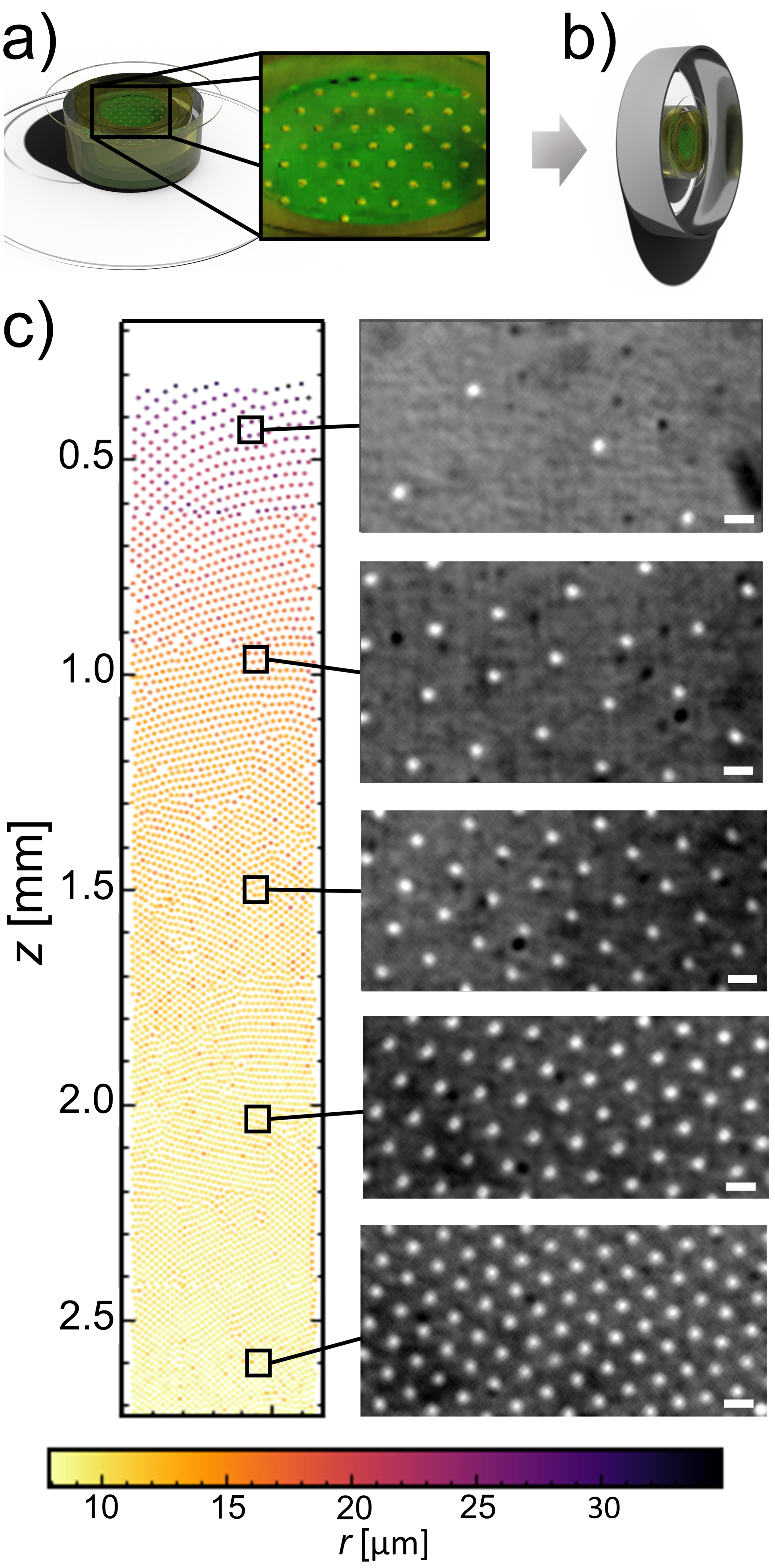}
    \caption{\textbf{Illustration of the experiments.} (a) Scheme of the cell holding the water/hexadecane interface, with a hexagonally-packed colloidal monolayer. (b) After monolayer formation, the cell is flipped by 90$^\circ$ and allowed to equilibrate for $>$ 8 hrs. (c) Representative plot of the particle position as a function of height ($z$) in a vertical monolayer of PS particles ($d = 2.80 \,\mu$m) after 15 hrs, and corresponding dark field microscopy images. The uppermost particle is assigned $z$=0 mm, and particles are color-coded according to their average inter-particle distance $r$. Scale bars: 10 $\mu$m.
    }
    \label{fig:Setup}
\end{figure}

The interactions among colloidal particles confined at a fluid interface are governed by a number of forces \cite{Oettel2008, Horozov2005, Garbin2012}, some of which are exclusive to the interface itself (\textit{e.g.}, capillary forces) \cite{Kralchevsky2000,Vialetto2022}. Concerning electrostatic forces, several studies have shown that the particles may exhibit strong, long-range repulsive interactions that can be orders of magnitude larger than those attained in a single fluid phase \cite{Aveyard2000,Horozov2003, Horozov2005, Pieranski1980}, making these colloidal monolayers crystallize at very low packing fractions \cite{Bonales2011}. Such repulsion arises from an asymmetric dissociation of the charged groups on the particle surface in the two fluids and from the material discontinuity, \textit{i.e.} different dielectric properties, across the interface. In addition to the classic DLVO screened-Coulomb term, a multipolar expansion of the field becomes necessary to fully describe the electrostatic interactions \cite{Hurd1985}. Particularly intriguing is the case of spherical hydrophobic charged colloids adsorbed at a water/non-polar fluid interface. In this case, several experimental techniques, (\textit{e.g.} optical tweezers \cite{Park2008, Park2010, Park2011, Park2014, Masschaele2010, Aveyard2002}, microstructural investigations \cite{Parolini2015, Masschaele2010}, collective sinking of particle monolayers \cite{Lee2017}, and compression in a Langmuir-Blodgett trough \cite{Petkov2014, Aveyard2000}) have repeatedly confirmed that a pair-wise dipolar repulsion of the form $U(r) \propto 1/r^3$ is the dominant term in the electrostatic interaction potential $U$ as a function of inter-particle distance $r$.  

In spite of the agreement on the functional form of the potential, different mechanisms have been proposed to describe the microscopic origins of the electrostatic dipoles, with the two main ones invoking either charge asymmetry on the aqueous side of the interface \cite{Hurd1985, Paunov2003, Park2008, Oettel2008}, or the presence of residual charges exposed to the oil \cite{Aveyard2002, Danov2006}, which remain essentially unscreened due to the very large Debye length in the non-polar phase. Surprisingly, even though for many polystyrene latexes and hydrophobized silica particles the latter explanation seems to be more likely, the only attempts to tune the interaction potentials have been through the use of additives (\textit{e.g.}, salts, surfactants) to the aqueous sub-phase \cite{Masschaele2010, Wirth2014, Horozov2003}, or via surface modifications to change particle wettability by means of silanization \cite{Horozov2003, Horozov2005}. Conversely, electrostatic interactions in bulk non-polar solvents \cite{Hsu2005, Smith2015} and in proximity of fluid interfaces \cite{Leunissen2007} have been extensively investigated together with ways to regulate them by adding organic salts or charge-regulating surfactants \cite{Yethiraj2003, Hsu2005}.   

Inspired by those studies and addressing an opportunity to fill an apparent gap in the literature, in this manuscript we show that the strength of the dipolar inter-particle interaction potential at water/oil interfaces can be continuously tuned over orders of magnitude upon introducing minor amounts of a charge-regulating organic salt, the ionic liquid Trihexyltetradecylphosphonium decanoate, into the non-polar phase. We extract the interaction potential by measuring the inter-particle distance as a function of height in a vertical monolayer under the action of gravity \cite{Horozov2003, Horozov2005, Law2011,Sakka2014}. This method, compared to the ones mentioned above offers the advantage of a robust, reproducible and statistically-relevant measurement over thousands of particles in a simple optical setup. 

We first create a macroscopically flat, horizontal water/hexadecane interface inside an experimental cell made out of two concentric glass rings glued on a glass cover slip. A schematic of the experimental cell is shown in Fig. S1, and further details of the experimental setup can be found in \cite{SI}. The inner and outer glass rings of the experimental cell have an inner diameter of 6 and 14 mm, and a height of 3 and 5 mm, respectively. The inner glass ring is filled with the aqueous phase until the surface is pinned to the edge (84 $\mu$L). Hexadecane is then pipetted on top to fill up the outer glass ring. In order to minimize the presence of surface-active contaminants, the hexadecane was preventively purified by three times extraction through an alumina and silica gel column \cite{SI}. 

We used different sulfonated polystyrene (PS) particles, with respective diameter and zeta potential $d = 2.80 \,\mu$m and $\zeta = -45.2 \pm 1.1$ mV, $d = 2.48 \,\mu$m and $\zeta = -45.3 \pm 1.2$ mV, and $d = 2.07 \,\mu$m and $\zeta = -35.2 \pm 1.8$ mV, and obtained 2D, hexagonally-packed monolayers (lattice spacing $\approx$7$D$) by spreading 0.5 $\mu$L of a 3:1 surfactant-free particle suspension:isopropanol mixture directly at the interface (Fig. \ref{fig:Setup}a). After sealing, the cell was carefully flipped by 90$^\circ$, bringing the particle-loaded interface into a vertical position, and was allowed to equilibrate for at least 8 hrs (Fig. \ref{fig:Setup}b). Under gravity, a gradient in inter-particle lattice spacing ($r$) is established as a function of height ($z$). We imaged this gradient by localizing particles from up to 18 stitched consecutive dark-field images of 300x1000 $\mu$m$^2$, see Fig. \ref{fig:Setup}c.

Previous reports to extract the inter-particle interaction potential from the barometric density profile of a vertical particle monolayer required the numerical integration of $(U)r$ \textit{vs.} $z$  \cite{Sakka2014}. However, this approach suffers from potential errors arising from the size of the integration step and several assumptions have to be made on system parameters, such as the contact angle and the packing arrangement of the particles, which contribute to another degree of variability at every integration step. Therefore, we instead directly fit the raw experimental $z$ versus $r$ data to measure the potential. Assuming the well-established form of the repulsive dipolar potential ($U$) at the fluid interface to follow \cite{Hurd1985}: 
\begin{equation}
    \frac{U(r)}{k_B T} = a_2 \frac{1}{r^3}
\end{equation}
it is straightforward to derive that the raw $z$ \textit{vs.} $r$ data must obey \cite{SI}: 
\begin{equation}\label{eq:zvsr}
    z(r) = A_2 \frac{1}{r^3} + C_2
\end{equation}
where $C_2$ is an integration constant, and $A_2$ relates to the prefactor $a_2$ commonly used in literature \cite{Park2010, Park2011, Park2014, Wirth2014, Masschaele2010} as: 
\begin{equation}\label{eq:a_2}
    a_2 = \frac{A_2 m^* g}{k_B T}
\end{equation}
where $g$ is the gravitational constant, $k_B$ is the Boltzmann constant, and $T$ the absolute temperature. 

\begin{figure}
    \centering
    \includegraphics[width=\columnwidth]{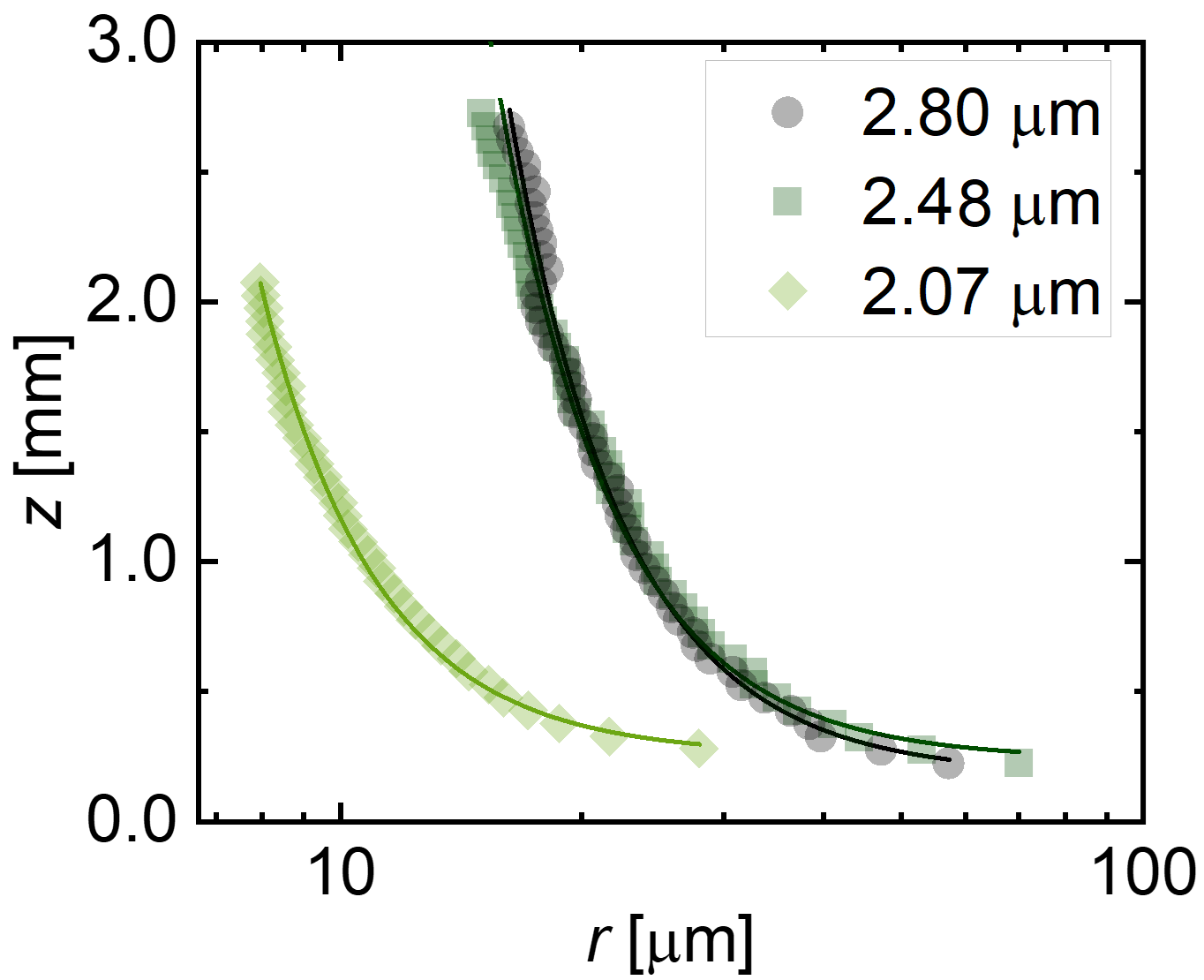}
    \caption{\textbf{Particle vertical position $z$ versus inter-particle distance $r$.} Data for three different sets of PS particles. Solid lines represent the fits to eq. \ref{eq:zvsr}.
    }
    \label{fig:207vs280}
\end{figure}

Figure \ref{fig:207vs280} shows a plot of $z$ \textit{vs.} $r$ for the different PS particles we used. Starting with the $d = 2.80 \,\mu$m particles (gray circles), using eq. \ref{eq:zvsr} we find an amplitude $A_2 = 7.9 \cdot 10^{6} \,\mu \mbox{m}^4$, corresponding to a value of $a_2 = 4.9 \cdot 10^{-12} \mbox{m}^3$. This value is on the higher end of the literature values, \textit{i.e.}, ranging from $5.0 \cdot 10^{-14}$ to $1.2 \cdot 10^{-12} \mbox{m}^3$, as observed by other groups using optical tweezers \cite{Park2011, Park2014, Wirth2014}, or a combination of optical tweezers and other analysis techniques \cite{Masschaele2010}. Notably, we calculate $a_2$ from multi-body interactions acting in the monolayer, and in the absence of any assumption on system parameters. The same procedure applied to the other two samples shows that the $d = 2.48 \,\mu$m particles behave very similarly to the previous ones (dark green squares), in line with the fact that they have a similar size, zeta potential and contact angle (see Fig. S2), while the smaller $d = 2.07 \,\mu$m particles (light green diamonds) retain the same $z\propto1/r^3$ dependence, albeit with $a_2 = 5.8 \cdot 10^{-13} \mbox{m}^3$, which is well within the range reported by the previously-mentioned works. 

Having bench-marked the interactions for the native particle systems, we move on to investigating the effect of additives to tune the electrostatic potential. In agreement with the literature \cite{Aveyard2002, Wirth2014, Sakka2014, Frydel2007}, we detected only a minor effect upon addition of ionic species (sodium chloride, NaCl) to the aqueous sub-phase (Fig. S5). The addition of up to 0.3M of NaCl did not modify the shape of the interaction potential as a function of $z$, nor did it affect the microstructure of the monolayer.

\begin{figure}
    \centering
    \includegraphics[width=\columnwidth]{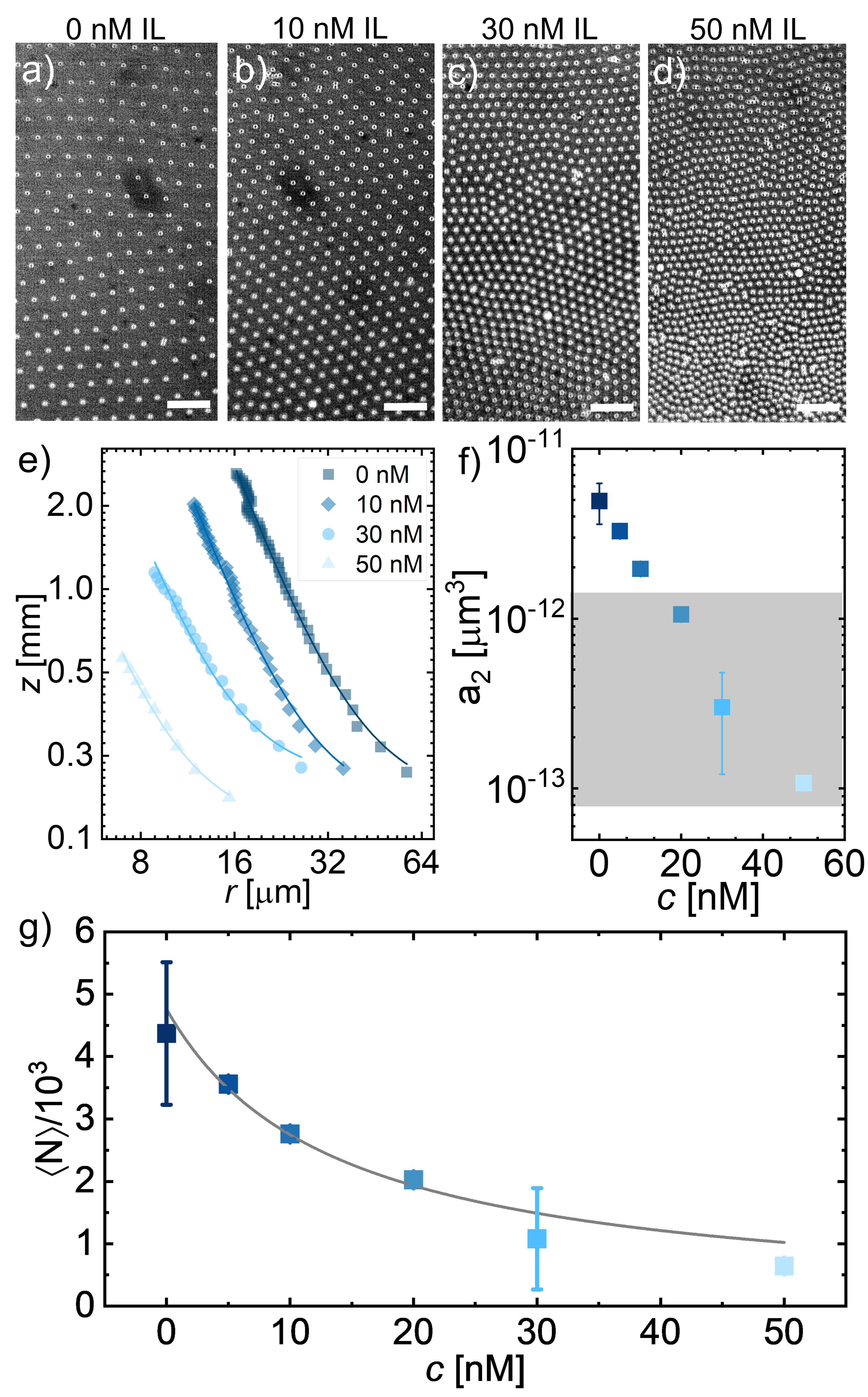}
    \caption{\textbf{The effect of adding IL to the organic phase.} 
    (a-d) Representative dark-field images of the  $d = 2.80 \,\mu$m PS monolayers at the bottom of the experimental cell (i.e. $z \approx$ 0.1-0.3 mm) for different IL concentrations. Scale bars: 20 $\mu$m. e) $z$ \textit{vs.} $r$ as a function of IL concentration. g) $a_2$ as a function of the concentration of IL $c$. The grey area marks the range of values reported in literature \cite{Park2010, Park2011, Park2014, Wirth2014, Masschaele2010}. (g) Average number of charges at the particle/oil interface $\langle N \rangle$ as a function of $c$. Line indicates the fit to eq. \ref{eq:AdsorptionIsoterm}.  
    }
    \label{fig:IL}
\end{figure}

Conversely, adding only trace amounts of an ionic liquid (Trihexyltetradecylphosphonium decanoate, IL) to the hexadecane had a striking effect on the inter-particle interaction potential and the resulting structure of the monolayers (Fig. \ref{fig:IL}a-d).  
The plots of $z$ \textit{vs.} $r$ shown in Fig. \ref{fig:IL}e reveal a shift towards smaller inter-particle distances with increasing amount of IL at the same $z$, while retaining an overall similar functional form, indicating that the magnitude of the dipolar interations is strongly affected by progressively introducing minute amounts of the IL into the organic phase. The values of $a_2$, plotted as a function of the concentration of ionic liquid $c$ (Fig. \ref{fig:IL}f) clearly show that the strength of the interaction potential decreases over roughly two orders of magnitude by increasing $c$ from 0 to 50 nM (Fig. \ref{fig:IL}). At $c$ = 50 nM, small aggregates in the denser region of the particle monolayer start to appear (Fig. \ref{fig:IL}d), while  for $c >$ 50 nM all particles aggregate into fractal-like structures (Fig. S5). This indicates that, under gravity and beyond 50 nM IL, the repulsive component of the potential reaches a magnitude that is comparable to the the attractive capillary force arising from out-of-plane undulations of the contact lines on the particles \cite{Horozov2005}, causing the particles to aggregate. Setting our extracted repulsive dipolar potential at 50 nM IL equal to the attractive capillary potential \cite{Stamou2000, Kralchevsky2001, Kralchevsky2016} (\textit{i.e.} $\frac{a_2 k_B T}{r^3} = 12 \pi \gamma H^2 \frac{R sin(\theta)^4}{r^4}$ with $R=d/2$ and the interfacial tension $\gamma$) for a minimum distance observed before aggregation of $r \approx 7 \mu m$, we estimate an amplitude of the contact angle undulations $H \approx$ 35 nm, which is comparable to literature values \cite{Kralchevsky2016}.

Previous work by Danov, Kralchevsky and others \cite{Danov2006, Kralchevsky2016} quantified the electrostatic dipolar interaction energy between two floating particles at a fluid-fluid interface as:
\begin{equation}\label{eq:dip_oil}
    U(r) = \frac{p_d^2}{2 \epsilon_0 \epsilon_n r^3}
\end{equation}
where, $\epsilon_n$ is  the  dielectric  constant  of  the non-polar phase, and $\epsilon_0$ is the permittivity of vacuum. The parameter $p_d$ is the effective dipole moment of the particle, which is $p_d = 4 \pi \sigma D(\theta, \epsilon_{pn}) R^3 sin^3(\theta)$, with $\sigma$ being the surface charge density at the particle/non-polar fluid interface, and $D(\theta, \epsilon_{pn})$ a tabulated dimensionless function \cite{Danov2006} that depends on the contact angle of the particles $\theta$, as well as the ratio of the dielectric constants of the particle and non-polar fluid $\epsilon_{pn}$. 

Considering the minute amounts of added IL, it is reasonable to assume that $\epsilon_n$ remains unchanged, and that screening remains insignificant ($\kappa R <<$ 1). Furthermore, we do not find any changes in $\theta$ due to the addition of IL (Fig. S2), suggesting that the modulation of the interaction potential directly results from changes in $\sigma$. Analogously, but opposite to the case charge regulation observed in systems of charged colloids in non-polar solvents \cite{Shapran2006, Royall2006, Sainis2008, Everts2016}, we assume our system to obey an associative discharging mechanism in which a single charged group on the particle surface $S^-$ can be occupied by a positive IL ion $P^+$. In this frame of reference, our system is expected to obey Langmuir-type adsorption that relates the average number of unoccupied effective charges at the particle/oil interface $\langle N \rangle$ to the total number of sites available for adsorption $\langle N \rangle_0$ \cite{Ninham1971}: 
\begin{equation}\label{eq:AdsorptionIsoterm}
    \langle N \rangle = \frac{\langle N \rangle_0}{1+K/c}
\end{equation}
where $K$ is a constant that depends on the equilibrium constant of the reaction $S^- + P^+ \rightleftharpoons SP$ and the surface potential. For our particles, we find $D(\theta, \epsilon_{pn})$ $\approx 1.05$ \cite{Danov2006}. Substituting the latter into eq. \ref{eq:dip_oil}, we are able to calculate $\sigma$, from which we directly derive the average number of charged groups per particle exposed to the oil $\langle N \rangle$. Fig. \ref{fig:IL}g shows the calculated values of $\langle N \rangle$ as a function of $c$. Indeed, we find that our data are well described by eq. \ref{eq:AdsorptionIsoterm}, as indicated by the fitted line in Fig. \ref{fig:IL}g.

In conclusion, we have shown that introducing nanomolar amounts of an organic salt into the non-polar phase allows for a continuous control over the strength of the repulsive dipolar interaction potential between charged polystyrene spheres confined at a water/oil interface. 
Our results furthermore strongly suggest that charge regulation by means of an associative discharging mechanism is responsible for the continuous control obtained over the strength of the interaction potential upon adding the IL.  
These findings underline that the presence of even ultra-low amounts of oil-soluble impurities might affect the measured interactions between adsorbed particles at fluid interfaces. They moreover offer an alternative perspective into which factors can be used to tailor the assembly of charged particles at a water/non-polar fluid interface, with interesting implications for processes such as emulsion destabilization or for fundamental studies on 2D crystallization.

\section*{Declaration of Competing Interest}
The authors declare that they have no known competing financial interests or personal relationships that could have appeared to influence the work reported in this paper.

\section*{Author Contribution Statement}
Author contributions are defined based on the CRediT (Contributor Roles Taxonomy) and listed alphabetically. Conceptualization: C.v.B., L.I., J.V. ; Formal analysis: C.v.B., L.I. ; Funding acquisition: L.I., J.V.; Investigation: C.v.B. ; Methodology: C.v.B., L.I., J.V. ; Project administration: L.I., J.V. ; Software: C.v.B. ; Supervision: L.I. ; Validation: C.v.B. ; Visualization: C.v.B., L.I., J.V. ; Writing – original draft: C.v.B., L.I., J.V. ; Writing – review and editing: C.v.B., L.I., J.V..

\section*{Acknowledgements}
The authors thank Svetoslav Anachov, Eric Dufresne, Jan Vermant, Martin Oettel and Alvaro Dominuguez for inspiring discussions and Federico Paratore for support. C.v.B. acknowledges funding from the European Union’s Horizon 2020 MSCA-ITN-ETN, project number 812780. J.V. acknowledges funding from the European Union’s Horizon 2020 research and innovation programme under the Marie Skłodowska Curie grant agreement 888076.

\FloatBarrier

\bibliographystyle{achemso}
\bibliography{bibliography}

\end{document}


\title{Supplementary Information for: \\
Tuning electrostatic interactions of colloidal particles at oil-water interfaces with organic salts}

\date{\vspace{-5ex}}

\author{Carolina van Baalen}
\author{Jacopo Vialetto%
     \thanks{Current address: Department of Chemistry, University of Florence, via della Lastruccia 3, Sesto Fiorentino, I-50019 Firenze, Italy}}
\author{Lucio Isa%
    \thanks{E-mail: lucio.isa@mat.ethz.ch}}
\affil{Laboratory for Soft Materials and Interfaces, Department of Materials, ETH Z{\"u}rich, Vladimir-Prelog-Weg 5, 8093 Z{\"u}rich, Switzerland}

\maketitle

\subsection*{SI1. Experimental setup and solvent purification}

\begin{figure}[h!]
    \renewcommand{\thefigure}{S1}
    \centering
    \includegraphics[width=\textwidth]{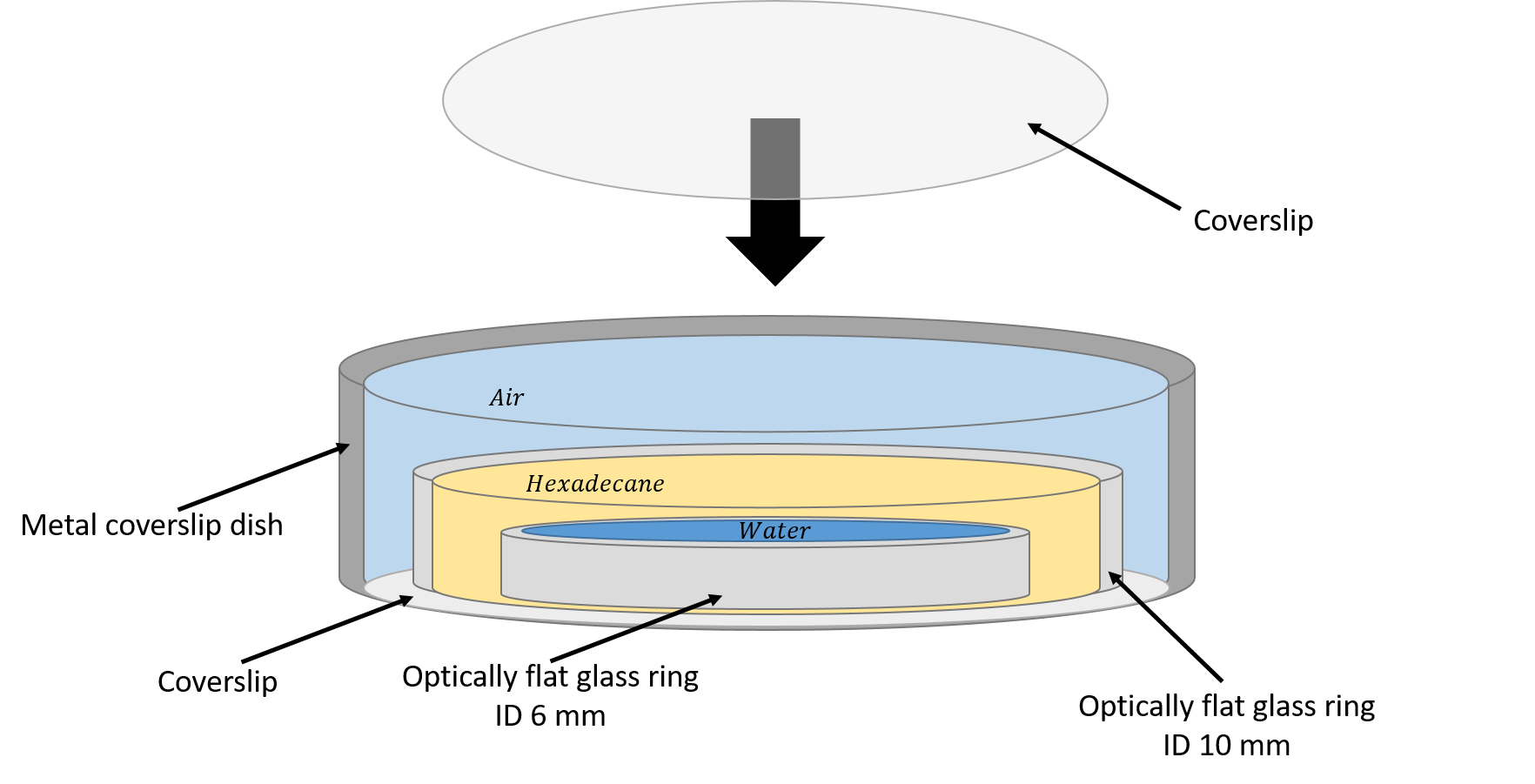}
    \caption{Schematic illustration of the cell. 
    }
    \label{fig:CellSchematic}
\end{figure}

Figure \ref{fig:CellSchematic} shows the cell used to create a macroscopically flat hexadecane-water interface on which the colloids are loaded following the procedure described in the main text. The experimental cell is made out of two concentric glass rings glued on a glass cover slip with a minimal amount of UV curable glue (NOA 63 Optical Adhesive). After depositing the upper coverslip on the outer glass ring, the ensemble is held together by capillary forces, which allow for flipping the cell without spilling the liquid.

The purification process of the hexadecane (AcrosOrganics) comprised three extractions through an alumina (EcoChromTM, MPAluminaB Act.1) and silica gel 60 (Merck) column, followed by measuring the interfacial tension of a drop of water in hexadecane, using a Drop Shape Analyzer system (KR\"{U}SS DSA100E). After the third extraction, the interfacial tension increased from 48 to 54 mN/m. Using a micro syringe pipette with a flat PFTE tip (Hamilton, 701 N Micro SYR Pipette), the water/hexadecane interface was then loaded with charged polystyrene spheres (micorparticles GmbH) a solvent-assisted spreading technique, as described in the main text. 

The particles adsorbed at the vertical fluid interface were imaged by stitching consecutive dark-field images of 300x1000 $\mu$m$^2$, taken using an  upright microscope (Nikon Japan) flipped on the side, equipped with a long working distance objective (20X M Plan APO, Optem) and a CCD camera (Ximea MQOZLMG-CM).

\FloatBarrier

\subsection*{SI2. Measurement of the particle contact angle}

The contact angle of the different PS particles at the water-hexadecane interface was measured by the gel trapping technique \cite{Paunov2003}. The measurements were performed as follows: the particles were first spread at an interface between an aqueous solution of gellan gum (2 wt \%) and hexadecane kept at 80$^\circ$C. Subsequently, the samples were cooled down to room temperature, allowing the gel to set and immobilize the particle monolayer. The hexadecane was then carefully removed and replaced with UV curable glue (NOA 63 Optical Adhesive). After curing the glue for 30 min. under UV light, the glue was peeled off from the gel surface together with the particles. The particle protrusion from the interface ($h$) was then obtained by atomic force microscopy (AFM, Bruker Icon Dimension) in tapping mode. The particle contact angle ($\theta$) is inferred by the following relation:
\begin{equation}
\theta = \cos^{-1} (\frac{h}{R} - 1)
\end{equation} 

\begin{figure}[h!]
    \centering
    \renewcommand{\thefigure}{S2}
    \includegraphics[scale=.6]{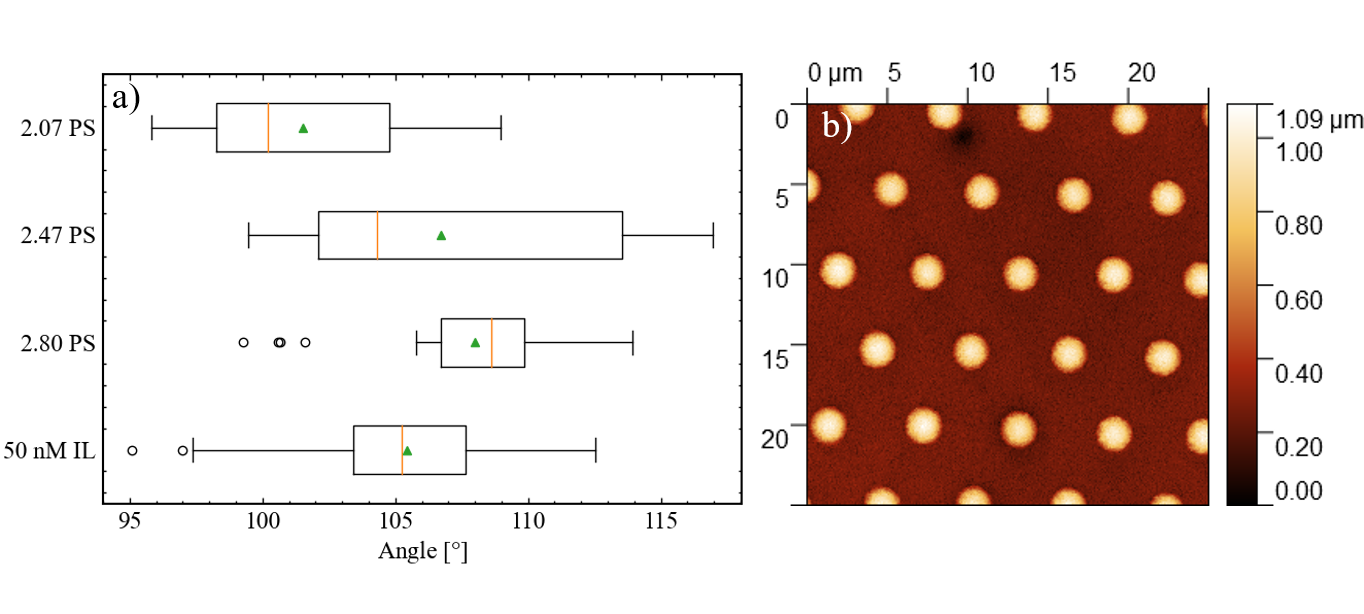}
    \caption{Contact angle of polystyrene particles at the water-hexadecane interface measured using the gel trapping technique. a) Box-plots of the contact angles measured for the three different sets of polystyrene particles with $D = 2.07$, $D = 2.48$, and $D = 2.80\,\mu$m with and without IL. Green triangles mark the averages. b) Typical AFM height image of the UV-curable glue containing the trapped particles from which the contact angle of the particles was extracted.
    }
    \label{fig:SI_GTT}
\end{figure}

\newpage

\subsection*{SI3. Evolution of the z vs. r profile over time}

\begin{figure}[h!]
    \centering
    \renewcommand{\thefigure}{S3}
    \includegraphics[width=0.7\textwidth]{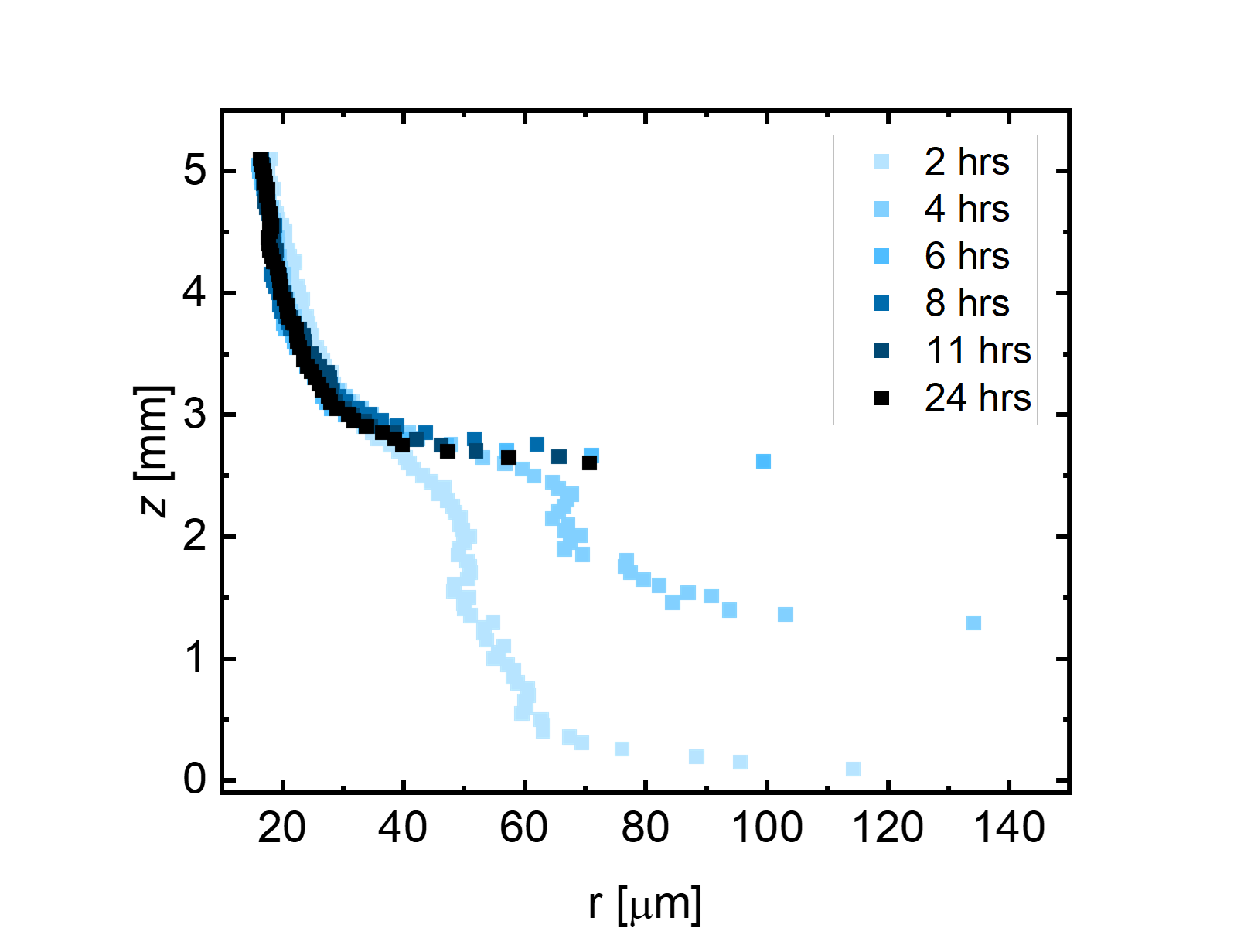}
    \caption{Time evolution of height $z$ vs. inter-particle distance $r$ measured for the PS particles with $D = 2.80 \,\mu$m confined at the water-hexadecane interface. For clarity the data has been rescaled to set the highest point in the particle monolayer after 2 hours to $z$ = 0 mm. 
    }
    \label{fig:TimeEvolution}
\end{figure}

\subsection*{SI4. Derivation of the fit to $z$ \textit{vs.} $r$}
The inter-particle interaction potential at the fluid interface can be obtained from the inter-particle distance $r$ \textit{vs.} height $z$ by numerically integrating the potential increase per particle resulting from the compression of the particle film using the following relation \cite{Sakka2014}:
\begin{equation}
    \mathrm{d}U = -P(r) \mathrm{d}A = -\sqrt{3} P(r)r \mathrm{d}r
\end{equation}
where $P$ is the 2D pressure, calculated by integrating the relation 
\begin{equation}
    \mathrm{d}P(z) = -\frac{2}{\sqrt{3}}\frac{1}{r^2}m^*g\mathrm{d}z
\end{equation}
with $g$ the gravitational acceleration, and the effective mass $m^*$, which can be calculated as
\begin{equation}\label{eq:effectivemass}
    m^* = \frac{4}{3}\pi R^3 (\rho_p-\rho_w) \left[ 1 + \frac{(\mathrm{cos^3}\theta-3\mathrm{cos}\theta + 2)(\rho_w-\rho_o)}{4(\rho_p-\rho_w)} \right]
\end{equation}
where $\rho_o$, $\rho_w$, and $\rho_p$ refer to the density of the oil, water and particles, respectively, and $R = D/2$.

Below we derive how the raw data $r$ \textit{vs.} $z$. can be used to obtain quantitative information about the potential starting from the established functional form of the dipolar potential \cite{Hurd1985}.

Take the derivative of potential $U$ with respect to $r$: 

\begin{equation}
    U = \frac{a_2}{r^3} \Rightarrow \mathrm{d}U = -3\frac{a_2}{r^4} \mathrm{d}r
\end{equation}

Set eq. 1 equal to eq. 2: 

\begin{equation}
    -3\frac{a_2}{r^4} \mathrm{d}r = -\sqrt{3} P(r)r \mathrm{d}r
\end{equation}

Rewrite: 

\begin{equation}
    \Rightarrow P(r) = \sqrt{3} \frac{a_2}{r^5}
\end{equation}
Take the derivative: 

\begin{equation}
    \frac{\mathrm{d}P(r)}{\mathrm{d}r} = -5 \sqrt{3}\frac{a_2}{r^6} 
\end{equation}
\\
\begin{equation}
    \frac{\mathrm{d}P(z)}{\mathrm{d}z} = \frac{\mathrm{d}P(r)}{\mathrm{d}r} \frac{\mathrm{d}r}{\mathrm{d}z}
\end{equation}
Insert eq. 4: 
\begin{equation}
    \frac{\mathrm{d}P(z)}{\mathrm{d}z} = -5 \sqrt{3}\frac{a_2}{r^6} \frac{\mathrm{d}r}{\mathrm{d}z}
\end{equation}

\begin{equation}
    \mathrm{d}P(z) = \frac{-2}{\sqrt{3}}\frac{1}{r^2} m^*g\mathrm{d}z
\end{equation}

\begin{equation}
    \frac{\mathrm{d}P(r)}{\mathrm{d}z} = \frac{-2}{\sqrt{3}}\frac{1}{r^2} m^*g = -5 \sqrt{3}\frac{a_2}{r^6} \frac{\mathrm{d}r}{\mathrm{d}z}
\end{equation}

\begin{equation}
    \mathrm{d}z = \frac{15}{2m^*g}\frac{a_2}{r^4} \mathrm{d}r
\end{equation}

Integrate: 

\begin{equation}
    z(r) = \frac{-5}{2m^*g}\frac{a_2}{r^3} + C2 = A_2\frac{1}{r^3} + C_2
    \label{eq:zvsr}
\end{equation}

\newpage

\subsection*{SI5. Effect of NaCl on the $z$ vs. $r$ profile}

\begin{figure}[h!]
    \renewcommand{\thefigure}{S4}
    \centering
    \includegraphics[width=0.7\textwidth]{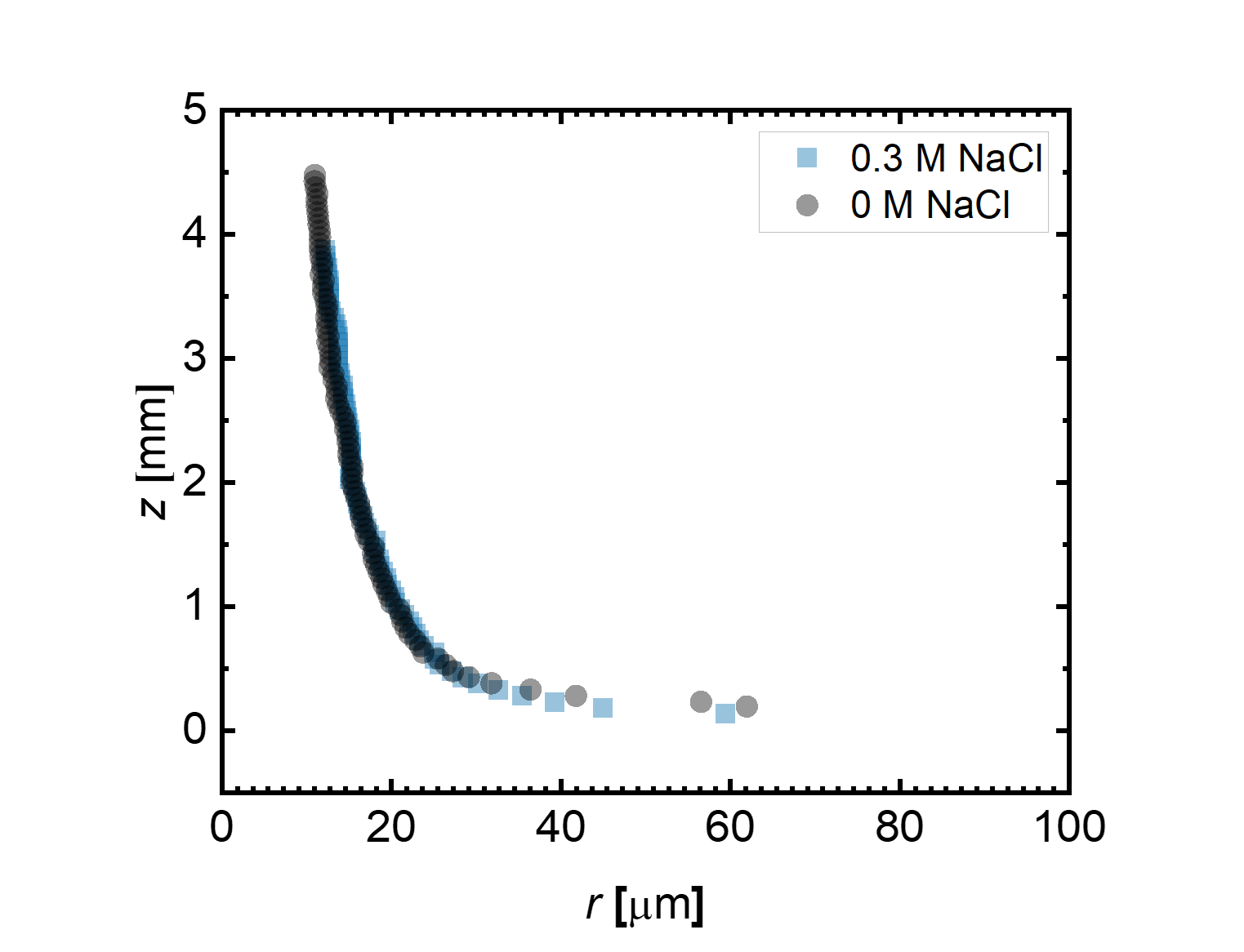}
    \caption{Plot of the height within the monolayer $z$, as a function of the inter-particle distance $r$, with and without 0.3 M NaCl. Data was extracted for PS particles with $D = 2.80\,\mu$m.
    }
    \label{fig:NaCl}
\end{figure}

\newpage

\subsection*{SI6. Effect of IL concentrations beyond 50 nM}

\begin{figure}[h!]
    \renewcommand{\thefigure}{S5}
    \centering
    \includegraphics[width=0.7\textwidth]{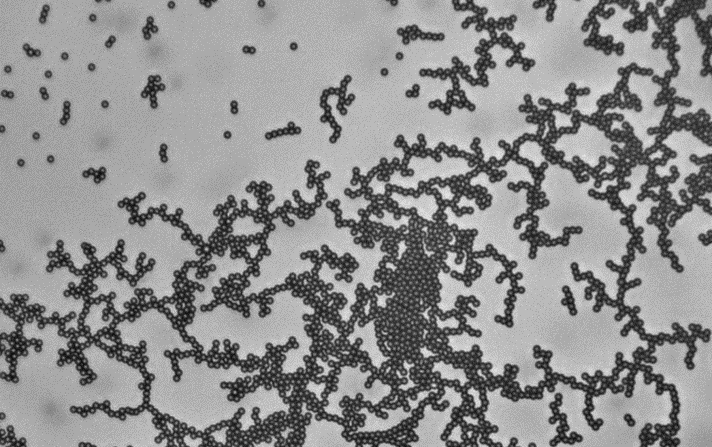}
    \caption{Representative microscopy image of a vertically placed monolayer of PS particles with $D = 2.80\,\mu$m at an hexadecane-water interface with 80 nM IL added to the hexadecane phase.
    }
    \label{fig:IL80nM}
\end{figure}

\FloatBarrier

\bibliographystyle{achemso}
\bibliography{bibliography}